\documentclass[conference]{IEEEtran}
\IEEEoverridecommandlockouts
\usepackage{cite}
\usepackage{amsmath,amssymb,amsfonts}
\usepackage{algorithmic}
\usepackage{graphicx}
\usepackage{textcomp}
\usepackage{xcolor}
\usepackage{comment}
\usepackage{footmisc}
\def\BibTeX{{\rm B\kern-.05em{\sc i\kern-.025em b}\kern-.08em
    T\kern-.1667em\lower.7ex\hbox{E}\kern-.125emX}}
\begin{document}

\title{Multi-Object Tracking for Collision Avoidance Using Multiple Cameras in Open RAN Networks  \\
\thanks{This work has been funded by the ”Ministerio de Asuntos Económicos y Transformación Digital” and the European Union-NextGenerationEU in the frameworks of the ”Plan de Recuperación, Transformación y Resiliencia” and of the ”Mecanismo de Recuperación y Resiliencia” under references TSI-063000-2021-18/24/77.}
}

\author{\IEEEauthorblockN{Jordi Serra}
\IEEEauthorblockA{\textit{CTTC/CERCA} \\
jserra@cttc.es}
\and
\IEEEauthorblockN{Anton Aguilar}
\IEEEauthorblockA{\textit{CTTC/CERCA} \\
aaguilar@cttc.es}
\and
\IEEEauthorblockN{Ebrahim Abu-Helalah}
\IEEEauthorblockA{\textit{CTTC/CERCA} \\
aebrahim@cttc.es}
\and
\IEEEauthorblockN{Ra\'{u}l Parada}
\IEEEauthorblockA{\textit{CTTC/CERCA} \\
rparada@cttc.es}
\and
\IEEEauthorblockN{Paolo Dini}
\IEEEauthorblockA{\textit{CTTC/CERCA} \\
pdini@cttc.es}}

\maketitle

\begin{abstract}
This paper deals with the multi-object detection and tracking problem, within the scope of open Radio Access Network (RAN), for collision avoidance in vehicular scenarios. To this end, a set of distributed intelligent agents collocated with cameras are considered. The fusion of detected objects is done at an edge service, considering Open RAN connectivity. Then, the edge service predicts the objects' trajectories for collision avoidance. Compared to the related work a more realistic Open RAN network is implemented and multiple cameras are used.
\end{abstract}


\section{Introduction}
The cornerstone of Intelligent Transportation Systems (ITS) is to provide safety to drivers and pedestrians. To achieve this aim, road agents, such as on-road cameras or vehicles equipped with cameras, share information to prevent hazards. Thereby, a low-latency connectivity is important to share road agents' information, thus preventing  collisions. This ingredient will be provided by 6G networks, which will support vertical services requiring ultra-reliable and low-latency communications such as ITS \cite{ITS_6G}. One of the key technologies of 6G will be Open RAN networks, as its openness and cloudified nature enables dynamic scaling and efficient management of network resources to support heterogeneous vertical services. 

Collision avoidance in ITS requires the detection and tracking of objects based on analyzing the video data from cameras deployed in the scenario. In complex scenarios, e.g. high density of vehicles, multiple cameras are needed to improve the performance, as different cameras may detect the same object from different perspectives. Such an approach is known as Multi-Object Multi-Camera Tracking (MOMCT) system. Evaluating vehicular applications such as collision avoidance is a costly and risky task. Thus, the use of simulators is convenient. In this regard, CARLA is an open source simulator for training and testing autonomous vehicles applications \cite{dosovitskiy2017carla}.

This papers deals with the MOMCT problem for collision avoidance in ITS. The connectivity is provided by the implementation of a realistic Open RAN network. The contributions of this paper are summarized next. This paper considers a realistic Open RAN network to provide the connectivity between the distributed agents and the edge service in the collision avoidance ITS system. On the contrary, related work relies on simulated networking solutions \cite{prathiba2021intelligent} or they implement partly a realistic network, e.g. \cite{zhang2020sdr} considers only the physical and medium access layers in C-V2X. Similar to \cite{PF} herein Particle Filters (PF) are leveraged for tracking detected objects. However, herein unlike \cite{PF} multiple cameras are used for multi-object tracking.   

\section{MOMCT demonstration setup in Open RAN} 
\begin{figure}
 \centering
 \includegraphics[width=0.9\linewidth]{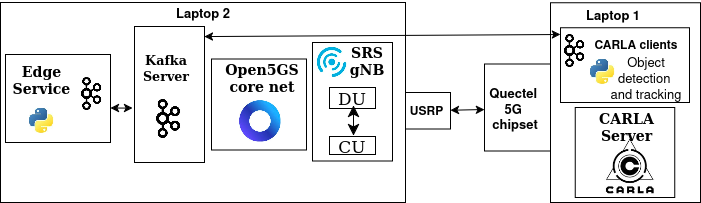} 
  \caption{Architecture of the MOMCT testbed in Open RAN}
 \label{fig:testbed_arch}
 \end{figure}

The demonstration consists of a set of distributed intelligent agents equipped with cameras,
within the context of CARLA simulator. Each intelligent agent detects objects by analyzing the video data of CARLA. Also, they extract a set of latitude and longitude points associated to the detected objects, a.k.a. tracklet. Those tracklets are sent to an edge service via the Open RAN network. Then, an edge service performs the fusion of tracklets from different cameras to predict the trajectory of the detected objects.
The architecture of the testbed for the MOMCT demonstration in Open RAN is shown in Fig. \ref{fig:testbed_arch}. In the next subsections the testbed functional blocks are explained along with its implementation.

\subsection{MOMCT functional blocks implementation}
\textbf{CARLA server}. It generates realistic vehicular environments, thus being a digital twin of vehicular scenarios. We have defined two cameras, a fixed camera in a traffic light and another outside a vehicle. Those cameras capture RGB video signal of the environment. Thus, the CARLA server provides video data and GPS data to the CARLA clients.

\textbf{CARLA clients}.  Perform object detection and tracking based on YOLOv8\footnote{https://yolov8.com}, by analyzing the video and GPS data generated by the cameras. Then, CARLA GPS and depth camera sensors are used for geo-location of objects and to generate the tracklets. A Kafka service is used to send the tracklets to the edge service via the Open RAN network. 

\textbf{Open RAN network}. The RAN is implemented by using SRS software\footnote{https://www.srslte.com/5g}, which supports the gNB disaggregation of Open RAN into CU, DU. The core network is implemented by using Open5Gs\footnote{https://github.com/open5gs/open5gs}.

\textbf{Edge Service}. It performs the fusion of incoming tracklets from different CARLA clients. Then, it predicts the trajectory for each detected object, which enables to estimate the risk of collision. The fusion and trajectory prediction are implemented by using PFs for each object. Each PF has a set of particles $\{\mathbf{x}^{i}\}_{i=1}^N$ with the expression $\mathbf{x}^{i} = \left[ x_{lat}, dx_{lat}, x_{lon}, dx_{lon}, w^i \right]^{T}$. Where $x_{lat}$,  $x_{lon}$, $dx_{lat}$, $dx_{lon}$, $w^i$ are the latitude, longitude, their speeds, and the particle's weight, respectively. The trajectory prediction is computed from the weighted average of the set of particles, i.e. $1/N \sum_{i=1}^{N}w^{i}\mathbf{x}^{i}$. See the paper gitlab\footnote{https://gitlab.cttc.es/supercom/momct-in-oran-demo-paper.git \label{github}} for further details. 

\subsection{MOMCT testbed in Open RAN}
The testbed implementation relies on the next hardware equipment that implements the functional blocks of the previous section. \textbf{Laptop 1}: equipped with an intel i5 processor, 16 GB RAM and a GPU Nvidia RTX3060 with 6GB. This laptop implements the CARLA client and CARLA server, which communicate via a socket.
The tracklets generated by CARLA client are transmitted via 5G to laptop 2, by using a Quectel RMU500EK 5G chipset. 
\textbf{Laptop 2}: it has an intel i9 processor, 128GB RAM and a GPU NVIDIA RTX A5500 with 16 GB. It is plugged to a USRP B210, which 
receives the 5G data. The USRP output is consumed by the Open RAN network.
On top of the network, there is a Kafka broker for the connectivity between the CARLA client and edge service. Finally, the edge service performs the fusion of tracklets and trajectories prediction.


\section{Experimental results}
\begin{figure}
 \centering
 \includegraphics[width=0.8\linewidth]{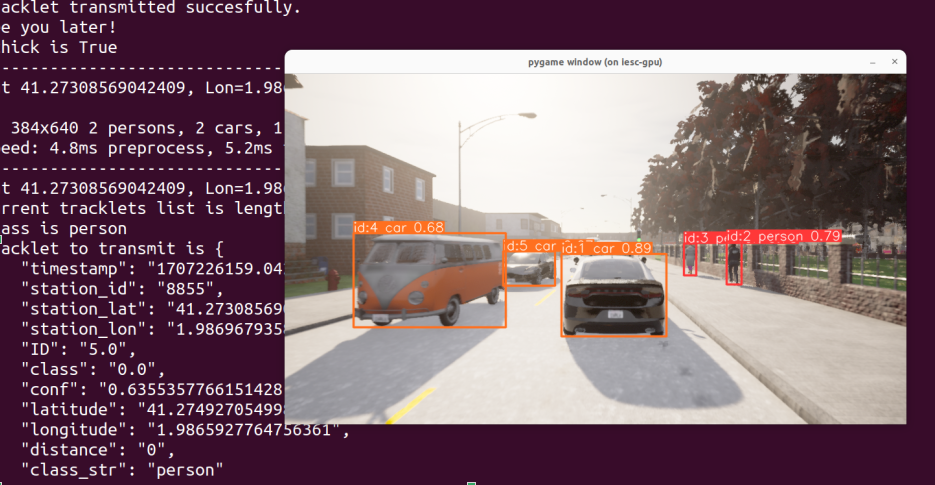}
 \caption{Object detection and tracking in CARLA clients}
 \label{fig:carla}
 \end{figure}

\begin{figure}
 \centering
 \includegraphics[width=0.7\linewidth]{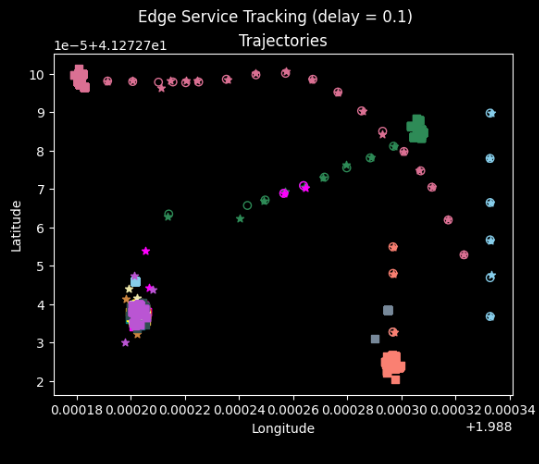}
 \caption{MOMCT edge service predicting trajectories}
 \label{fig:edge}
 \end{figure}

A screenshot of the CARLA client could be seen in Fig. \ref{fig:carla}. The client connects to CARLA server to obtain the streaming from the camera mounted on the connected vehicle. Client's Machine Learning (ML) model performs a classification on the frames to find relevant objects for the driving.
Using CARLA GPS sensor and depth camera, the client determines the geo-location of the detected objects in screen which is integrated to make a tracklet. Tracklets are sent to the edge service using the Open RAN network via Kafka. On the edge, tracklets from different cameras are asynchronously processed to aggregate them in trajectories. Fig. \ref{fig:edge} shows an example of this process. In the image, a camera mounted on a red light is used for the detection. The image shows 3 different data, the particles used for trajectory estimation (squares), the trajectory estimations (stars), and the ground truth (circles). Each color represents a  different object. In this case, lilac represents the camera, while the others are vehicles moving in front of it. One of them is a connected vehicle (pingk), although, the edge service is agnostic to the sources.
A video of the demonstration is provided in the paper gitlab\footref{github}.
 
\section{Conclusions}
This paper has presented a demonstration of the MOMCT problem for ITS in Open RAN. A set of distributed agents detect, track objects and send the information via Open RAN to an edge service. The latter performs the fusion of detected objects and trajectory prediction for collision avoidance. The implementation is based on the integration of a realistic Open RAN network and the CARLA vehicular simulator.

\vspace{12pt}

\end{document}